\documentclass[12pt]{iopart}
\usepackage{iopams} 
\usepackage{url}
\usepackage{float}
\usepackage{graphicx}
\usepackage{footmisc}

\def\units#1{\hbox{$\,{\rm #1}$}}
\def\degrees{\hbox{${}^\circ$}}
\newcommand{\vv}[1]{``{#1}''}

\begin{document}

\title[Measurement of angular correlation between $\gamma$ rays from $^{60}$Co]{Measurement of the angular correlation between the two gamma rays emitted in the radioactive decays of a $^{60}$Co source with two NaI(Tl) scintillators}

\author{
E.~C.~Amato$^1$, 
A.~Anelli$^1$,
M.~Barbieri$^1$,
D.~Cataldi$^1$,
V.~Cellamare$^1$,
D.~Cerasole$^1$,
F.~Conserva$^1$,
S.~De Gaetano$^{1,2}$,
D.~Depalo$^1$,
A.~Digennaro$^1$,
E.~Fiorente$^1$,
F.~Gargano$^2$,
D.~Gatti$^1$,
P.~Loizzo$^1$,
F.~Loparco$^{1,2,*}$,
O.~Mele$^1$,
N.~Nicassio$^1$,
G.~Perfetto$^1$,
R.~Pillera$^{1,2}$,
R.~Pirlo$^1$,
E.~Schygulla$^1$,
D.~Troiano$^1$
}

\address{$^1$ Dipartimento di Fisica \vv{M.~ Merlin} dell’Universit\`a e del Politecnico di Bari, I-70126 Bari, Italy}
\address{$^2$ Istituto Nazionale di Fisica Nucleare, Sezione di Bari, 70126 Bari, Italy}

\ead{\mailto{francesco.loparco@ba.infn.it}}

\begin{abstract}
We implemented a didactic experiment to study the angular correlation between the two gamma rays emitted in typical $^{60}$Co radioactive decays. We used two NaI(Tl) scintillators, already available in our laboratory, and a low-activity $^{60}$Co source. The detectors were mounted on two rails, with the source at their center. The first rail was fixed, while the second could be rotated around the source. We performed several measurements by changing the angle between the two scintillators in the range from $90\degrees$ to $180\degrees$. Dedicated background runs were also performed, removing the source from the experimental setup. We found that the signal rate increases  with  the angular  separation between the  two  scintillators,  with  small discrepancies from the theoretical expectations.
\end{abstract}

\section{Introduction}

Cobalt-60 ($^{60}$Co) is among the radioactive isotopes of cobalt with a spin-parity $J^{P}=5^{+}$ and undergoes beta decay with a half-life of $1925.20 \pm 0.25\units{days}$~\cite{1992NIMPA.312..349U,WOS:000173674600021}.
A $99.88\%$ fraction of the beta decays~\cite{LARA} leads to an excited state of $^{60}$Ni with a spin-parity $J^{P}=4^{+}$, which subsequently decays, passing through the intermediate state $J^{P}=2^{+}$, into the ground state $J^{P}=0^{+}$ with the emission of two gamma rays, respectively with energies of $1.17 \units{MeV}$ and $1.33 \units{MeV}$. 
Since the lifetime of the intermediate state is of the order of $10^{-12}$\units{s}~\cite{PhysRev.78.558} and is much smaller than typical experimental time resolutions, the two gamma rays are expected to be detected in coincidence.

General considerations of radiation theory show that the emission directions of consecutive gamma rays produced by an excited nucleus are correlated~\cite{Hamilton:1940zz}. These transitions involve three nuclear states and the multipole order of the emitted radiation determines the angular separations between the two photons. The angular correlation can be described in terms of the relative probability per unit solid angle $W(\theta)$
of the second photon to be emitted at an angle $\theta$ with respect to the first one. Explicit quantum mechanical calculations by Hamilton show that, apart from a constant factor, the correlation function has the general form:
\begin{equation}
    W(\theta) = 1 + \sum_{k=1}^{L} \alpha_{k} \left( \cos\theta \right)^{2k}
\end{equation}
where $L$ is the lowest angular momentum of the two gamma rays.
Calculations by Hamilton provide the theoretical values of the coefficients $\alpha_{k}$ in case of dipole and quadrupole radiation for all possible nuclear angular momenta. 
In the case of $^{60}$Co, both transitions can be assumed to be electric quadrupoles, and the angular correlation function is given by:
\begin{equation}
    W(\theta) = 1 + \frac{1}{8} \cos^{2}\theta + \frac{1}{24} \cos^{4}\theta
\end{equation}

Several measurements of the angular correlations between the photons produced in the $^{60}$Co decays were performed in the past and the experimental results were found in agreement with the predictions from Hamilton's model (see for instance refs.~\cite{PhysRev.78.558,PhysRev.70.978,Colombo:1950,Smith:2018amy,PhysRev.91.616}). We have designed and implemented a custom experimental setup to perform this measurement in a didactic laboratory for undergraduate Physics students with the instrumentation already available. 

\begin{figure}[!hb]
\centering
\includegraphics[width=0.95\columnwidth]{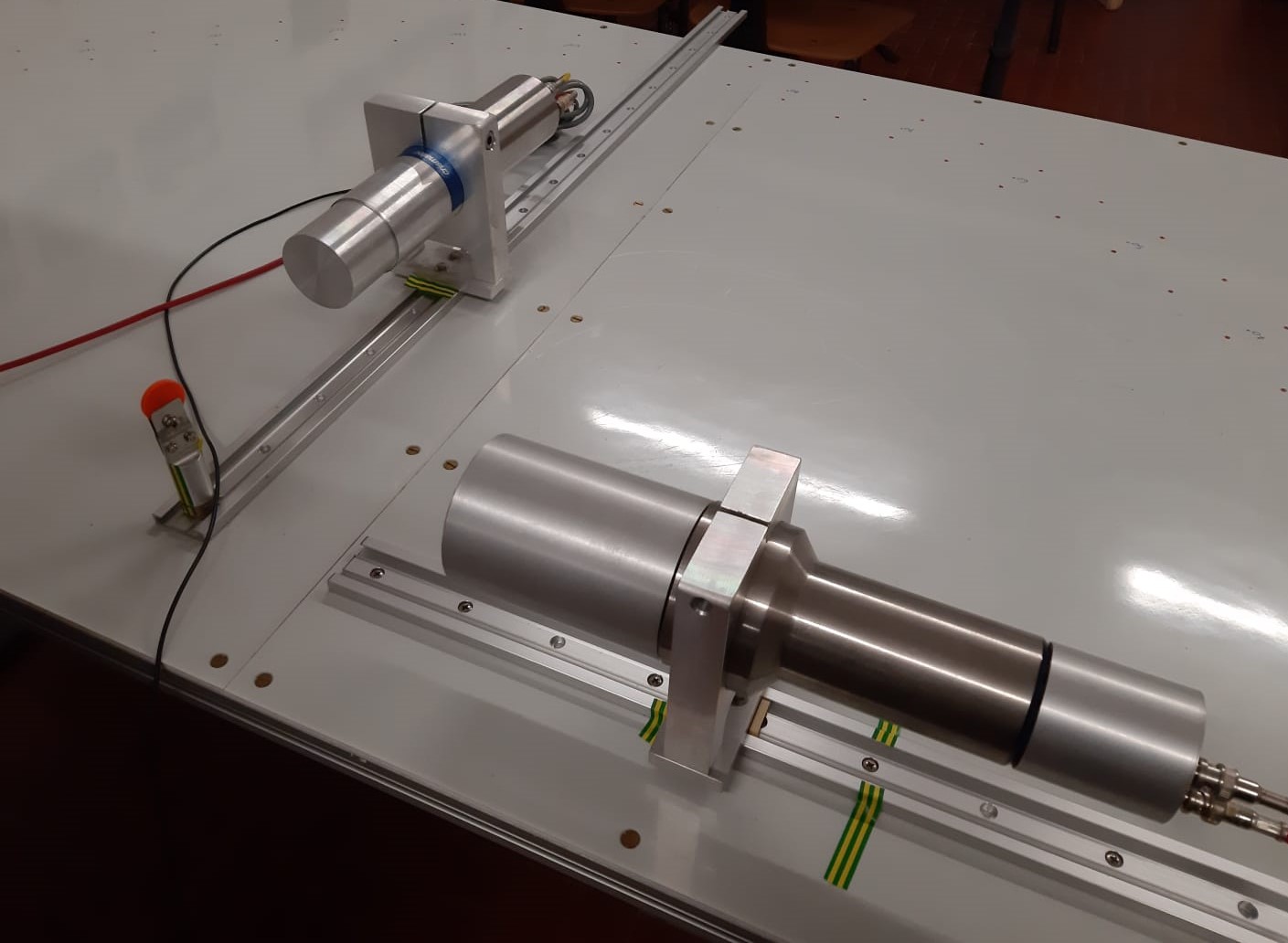}
\caption{Photo of the experimental setup. The scintillator $S_{0}$ is mounted on the fixed rail on the right, while the scintillator $S_{1}$ is mounted on the mobile rail on the left. The $^{60}$Co source is the orange disk placed in the holder shown at the center of the picture.}
\label{fig:setup}
\end{figure}

\section{Experimental setup}
\label{sec:setup}

\begin{figure}[!t]
\centering
\includegraphics [width=0.95\columnwidth]{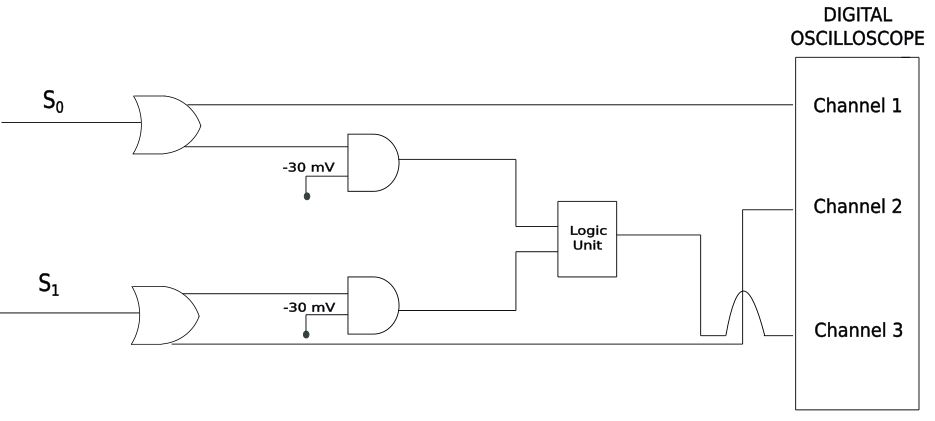}
\caption{Schematic of the trigger logic. The analog signals from $S_{0}$ and $S_{1}$  are sent to two fan-in/fan-out modules. Each module provides two copies of the corresponding input signal. A copy of the two signals is sent to the oscilloscope, while the other one is sent to a discriminator with a threshold fixed at $-30\units{ mV}$. The logic output signals from the  discriminators are sent to a logic unit, which provides the trigger signal to the oscilloscope.}
\label{fig:trigger}
\end{figure}

A photo of the experimental setup is shown in Fig.~\ref{fig:setup}. The $^{60}$Co source is a $1\units{mm}$ thick disk with a diameter of $1\units{cm}$ and an activity of $\sim 5 \units{kBq}$ placed on a holder mounted at the center of a table. The detectors are two cylindrical NaI(Tl) scintillators coupled with photomultiplier tubes (PMTs), which can be moved along two rails placed on the table. The first scintillator ($S_{0}$ hereafter), which has a diameter of $8.2 \units{cm}$ and a height of $8.2 \units{cm}$, is mounted on a fixed rail. The second scintillator ($S_{1}$ hereafter), which has a diameter of $5.8 \units{cm}$ and a height of $5.8 \units{cm}$, is mounted on a mobile rail, which can be rotated around the source with respect to the fixed one. A graduated scale, divided in sections of $5\degrees$ step between $0\degrees$ and $180\degrees$, is drawn on the table and is used to measure the angle between the two rails. Both scintillators can be also moved along the rails at different distances from the source. 

During our measurements, the scintillators $S_{0}$ and $S_{1}$ were placed at distances of $20 \units{cm}$ and of $14.1\units{cm}$ from the source, respectively. In this way, the solid angle subtended by each scintillator with respect to the source was $\Delta \Omega \simeq 0.132\units{sr}$. This geometry provided a good compromise between the need of taking data with a high enough rate and the need of keeping a limited uncertainty on the angle between the two scintillators. Indeed, the latter is determined by the radius of the two scintillators and by their distances from the source and it amounts to $\Delta \theta \sim 16 \degrees$.

The signals from the two scintillators were acquired using a digital oscilloscope \textit{Teledyne Le Croy WaveRunner 6 Zi}~\cite{OscilloscopeManual}. Fig.~\ref{fig:trigger} shows a scheme of the trigger logic. The analog signals from the PMTs coupled with $S_{0}$ and $S_{1}$ are sent to two linear fan-in/fan-out modules, producing two copies of each input signal. A copy of the two signals from $S_{0}$ and $S_{1}$ is sent to the oscilloscope (channels 1 and 2 in the scheme of Fig.~\ref{fig:trigger}). The second copy is sent to a discriminator module with a threshold set at $-30 \units{mV}$. The logic signals from the discriminator are sent to a logic unit, where the trigger to the oscilloscope is formed (channel 3 in the scheme of Fig.~\ref{fig:trigger}). 

The logic unit allows the implementation of different trigger configurations. We performed our measurements with the following ones:
\begin{itemize}
\item \vv{and} configuration: in this case we required both signals from $S_{0}$ and $S_{1}$ with pulse height exceeding the threshold. This configuration was implemented to select events with a gamma-ray in each scintillator;
\item \vv{or} configuration: in this case we required a signal above the threshold from either $S_{0}$ or $S_{1}$. This configuration was implemented to perform energy calibrations, as discussed in Sec.~\ref{sec:calibration};
\item \vv{single scintillator} configuration: in this case we required a signal above the threshold only from $S_{0}$ (or $S_{1}$). This configuration was also implemented to perform energy calibrations. 
\end{itemize}

\begin{figure}[!t]
\centering
\includegraphics [width=0.95\columnwidth]{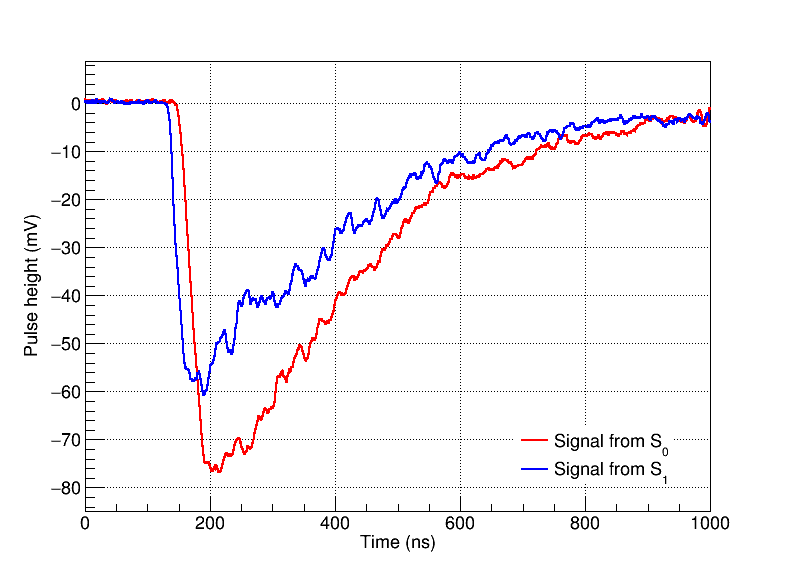}
\caption{Example of the analog signals produced in the two scintillators by the gamma rays emitted from the $^{60}$Co source.}
\label{fig:signals}
\end{figure}

Fig.~\ref{fig:signals} shows an example of the analog signals from the two scintillators recorded when the trigger logic was set in the \vv{and} configuration with the $^{60}$Co source. The signals are likely originated by two gamma rays from the source being absorbed in the two scintillators.
We see that both signals exhibit a rise time of a few tens of $\units{ns}$ and a fall time of a few hundreds of $\units{ns}$. These values are expected, since the decay time of the fluorescence light in NaI(Tl) scintillators is of $230\units{ns}$~\cite{knoll2010radiation} and the time constant $RC$ of the readout circuit is $\simeq 15 \units{ns}$ (this value is obtained from the oscilloscope input resistance  $R=50\units{\Omega}$ and from the cable capacitance of  about $300 \units{pF}$).

We used the oscilloscope to measure the amplitudes of the pulses from both scintillators and the integral of each pulse in a $1 \units{\mu s}$ time window. This information is important since the value of the time integral is  proportional to the charge released in each scintillator. We also measured the time intervals between events. The data taken were stored in ROOT files~\cite{ROOT}, which are easily accessible for data analysis. 

\newpage

\section{Energy calibration}
\label{sec:calibration}

\begin{figure}[!t]
\centering
\includegraphics[width=0.45\columnwidth]{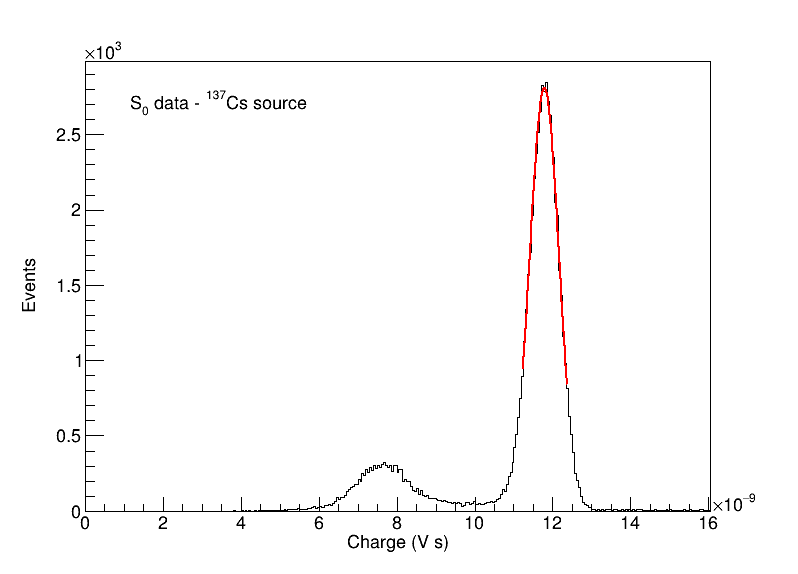}
\includegraphics[width=0.45\columnwidth]{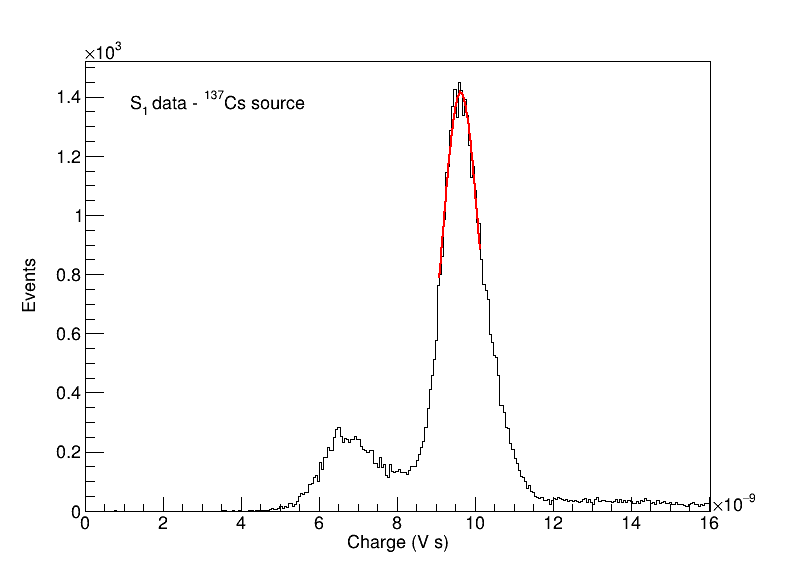}
\includegraphics[width=0.45\columnwidth]{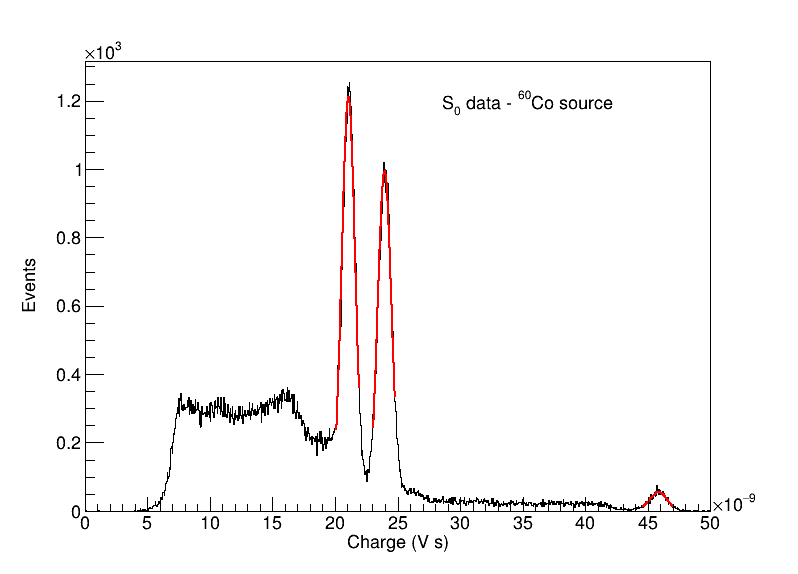}
\includegraphics[width=0.45\columnwidth]{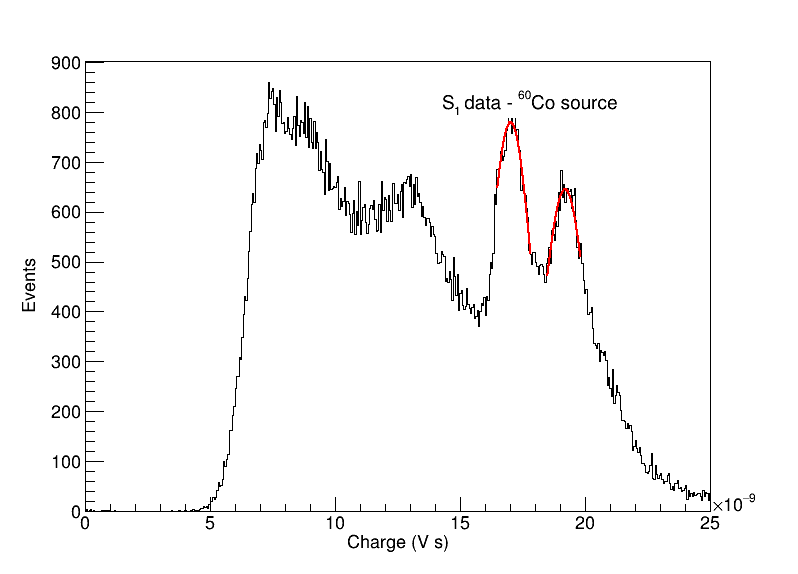}
\caption{Charge distributions obtained with the $^{137}$Cs source (top row)  and with the $^{60}$Co source (bottom row). The plots on left column refer to the scintillator $S_{0}$, while those on the right column refer to the scintillator $S_{1}$. All data have been taken in the \vv{single scintillator} trigger configuration. The full-energy peaks have been fitted with gaussian functions and the fits are indicated with red lines. In the case of $^{60}$Co source, the full-energy peak corresponding to the sum of the energies of the two photons in the scintillator $S_{1}$ was not clearly visible and was therefore not fitted.}
\label{fig:chargedistributions}
\end{figure}

We performed the energy calibration of the two scintillators with the $^{60}$Co source and with a $^{137}$Cs source. The $^{60}$Co source is used to calibrate the detectors with $1.17\units{MeV}$ and $1.33\units{MeV}$ photons. Moreover, if both photons are detected in the same scintillator, an additional calibration point corresponding to the sum of the energies of the two photons ($2.50\units{MeV})$ is available. The $^{137}$Cs source emits a gamma-ray of $0.662\units{MeV}$~\cite{LARA} and provides a calibration point at low energies.  

Fig.~\ref{fig:chargedistributions} shows the charge distributions obtained when taking data in the \vv{single scintillator} trigger configuration with the $^{60}$Co and $^{137}$Cs sources. The full-energy peaks in the two scintillators were fitted with gaussian functions. The charge values of each peak were evaluated as the mean values of the corresponding fit function. 

A further calibration point is provided by the position of the \vv{pedestal} peak, which corresponds to a null energy deposition in the scintillator. The position of the pedestal peak in each scintillator was evaluated from its charge distribution in a run with the $^{137}$Cs source in the \vv{single scintillator} trigger configuration, in which the trigger was provided by the other one. 

\begin{figure}[!t]
\centering
\includegraphics[width=0.95\columnwidth]{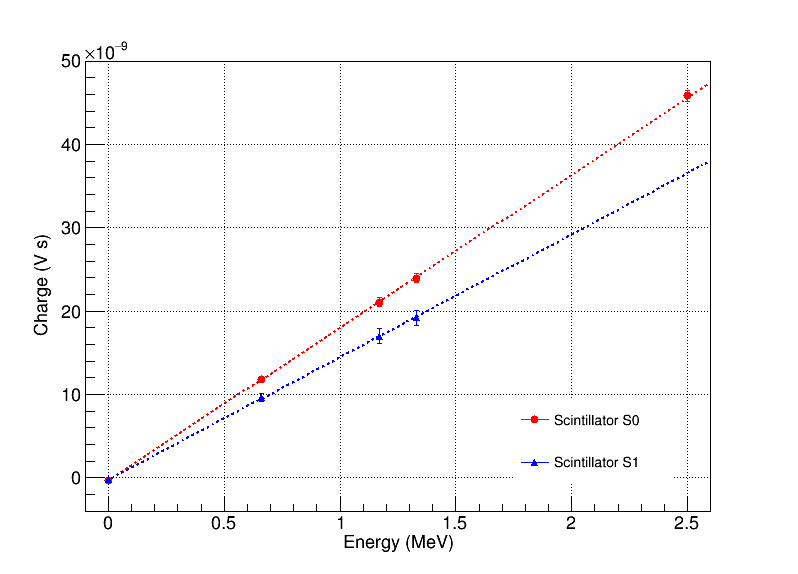}
\caption{Energy calibration of the two scintillators. The markers indicate the charges corresponding to the mean values of the gaussian functions fitting the peaks of the spectra obtained in the calibration runs. The data have been fitted with straight lines, which are also shown in the plot.}
\label{fig:calcurves}
\end{figure}

The results of the energy calibrations are summarized in Fig.~\ref{fig:calcurves}. The data of both scintillators are well fitted with straight lines. We see that the charge depends linearly on the energy deposited in the scintillators in the range of interest for our measurements.  

\begin{figure}[!t]
\centering
\includegraphics[width=0.95\columnwidth]{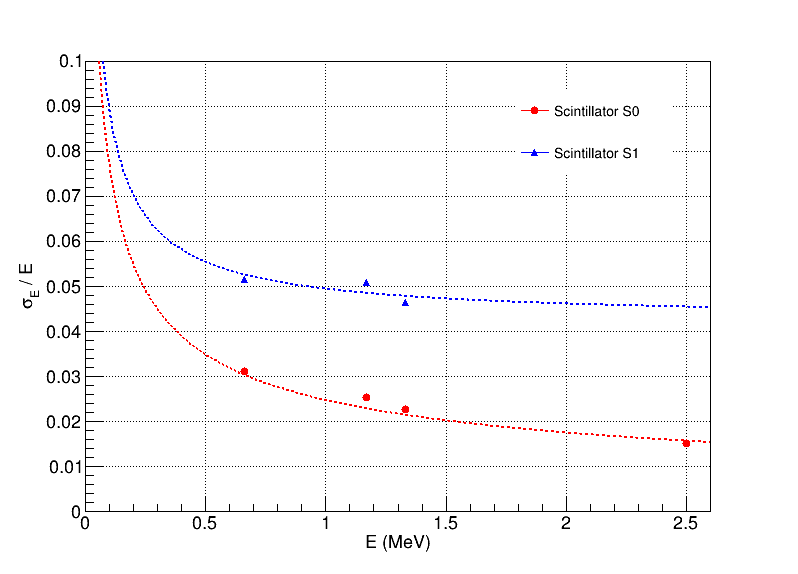}
\caption{Energy resolution as a function of the energy deposited in the scintillators. The data of $S_{0}$ and $S_{1}$ have been fitted with the functions in Eqs.~\ref{eq:resolution0} and ~\ref{eq:resolution1}, respectively. The best fit curves are represented by the dashed lines.}
\label{fig:resolution}
\end{figure}

Finally, we studied the behaviour of the energy resolution $\sigma_{E}/E$ as a function of the energy deposited in the scintillators. The energy resolution was evaluated from the fits of the full-energy peaks of the $^{60}$Co and $^{137}$Cs spectra used to perform the energy calibrations. The results
are shown in Fig.~\ref{fig:resolution}. We see that the energy resolution for the $^{60}$Co gamma rays is about $2.5\%$ for $S_{0}$ and about $5\%$ for $S_{1}$.

We also fitted our data with functions which are commonly used in high-energy physics to describe the energy resolution of electromagnetic calorimeters~\cite{Zyla:2020zbs}. In particular, the data of $S_{0}$ have been fitted with the function:
\begin{equation}
   \frac{\sigma_{E}}{E} = \frac{a}{\sqrt{E(\units{MeV})}}
\label{eq:resolution0}
\end{equation}
where $a=(2.464 \pm 0.006) \times 10^{-2}$, while the data of $S_{1}$
have been fitted with the function:
\begin{equation}
   \frac{\sigma_{E}}{E} = \frac{a}{\sqrt{E(\units{MeV})}} \oplus b
\label{eq:resolution1}
\end{equation}
where the $\oplus$ symbol indicates the sum in quadrature, $a=(2.508 \pm 0.181) \times 10^{-2}$ and $b=(4.254 \pm 0.095) \times 10^{-2}$. We see that the resolution of $S_{0}$ is mainly dominated by the stochastic term, while in the case of $S_{1}$ the constant term cannot be neglected.

\section{Event selection and data analysis}
\label{sec:dataanalysis}

To measure the angular correlations between the gamma rays emitted in the $^{60}$Co decays we performed several runs in the \vv{and} trigger configuration, changing the angle between the two detectors. Two sets of measurements were performed for each angle. The first measurement was performed with the $^{60}$Co source placed on its holder, while the second was performed removing the source from the experimental setup. The second set of measurements was needed to evaluate the background rates due to natural radioactivity and cosmic rays penetrating in our laboratory.  

\begin{figure}[!t]
\centering
\includegraphics[width=0.95\columnwidth]{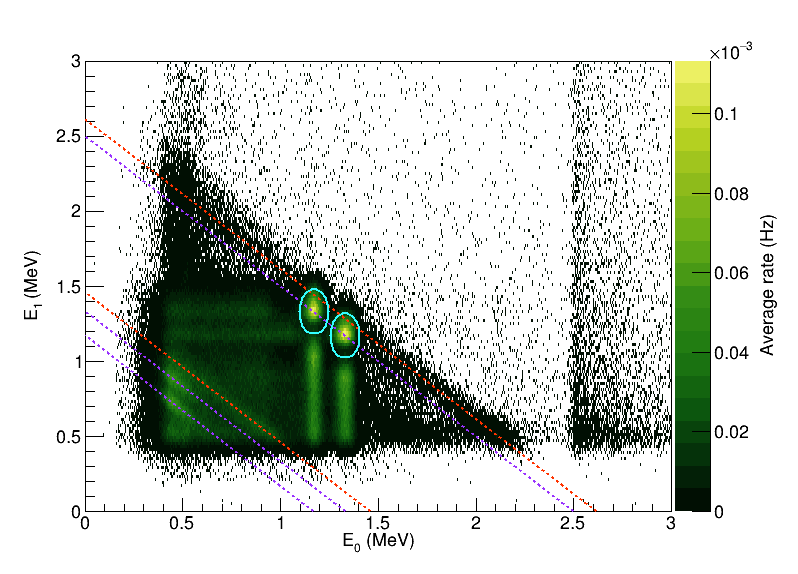}
\caption{Average rate of events as a function of the energy deposited in the two scintillators in the runs with the $^{60}$Co source. The violet dashed lines correspond to the energies of the two photons emitted by the $^{60}$Co source ($1.17 \units{MeV}$ and $1.33 \units{MeV}$) and to the sum of the energies of both photons ($2.50 \units{MeV})$. The orange dashed lines correspond to the energies of photons produced in the decays of $^{40}$K and $^{208}$Tl ($1.461\units{MeV}$ and $2.610\units{MeV}$, respectively). The cyan continuous lines indicate the ellipses used for the event selection.  
}
\label{fig:signal2dspectra}
\end{figure}

\begin{figure}[!t]
\centering
\includegraphics[width=0.95\columnwidth]{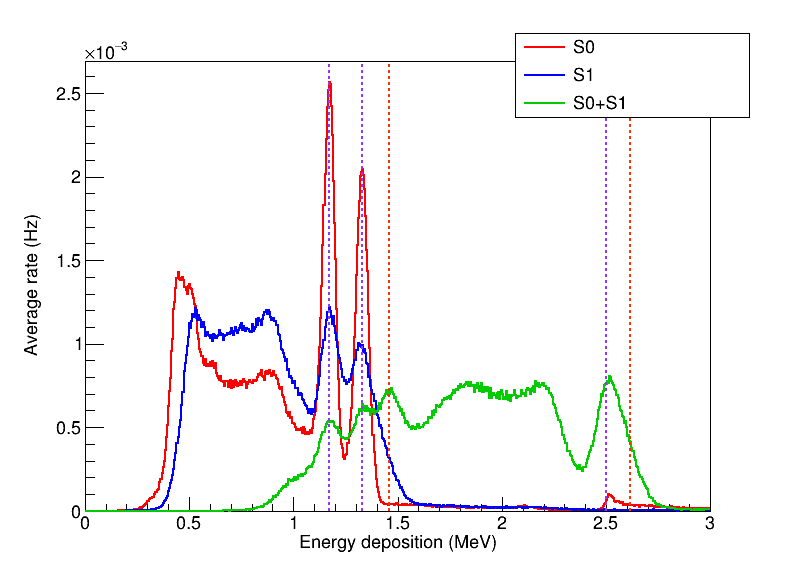}
\caption{Average rate of events as a function of the energy deposited in each of the two scintillators in the runs with the $^{60}$Co source. The violet dashed lines correspond to the energies of the two photons emitted by the $^{60}$Co source ($1.17 \units{MeV}$ and $1.33 \units{MeV}$) and to the sum of the energies of both photons ($2.50 \units{MeV})$. The orange dashed lines correspond to the energies of photons produced in the decays of $^{40}$K and $^{208}$Tl ($1.461\units{MeV}$ and $2.610\units{MeV}$, respectively).  
}
\label{fig:signal1dspectra}
\end{figure}

Figs.~\ref{fig:signal2dspectra} and ~\ref{fig:signal1dspectra} show a summary of the data taken by the two scintillators in the runs with the $^{60}$Co source. In Fig.~\ref{fig:signal2dspectra} we show the average rate of events as a function of the energies $E_{0}$ and $E_{1}$ measured by the two detectors. The following regions can be identified in the plot:
\begin{itemize}
\item the two peaks at the positions $(1.17\units{MeV},1.33\units{MeV})$ and $(1.33\units{MeV},1.17\units{MeV})$ correspond to events where one of the two photons from the $^{60}$Co decay is absorbed in $S_{0}$ and the second one is absorbed in $S_{1}$, with both gamma rays depositing their whole energy in the scintillator in which they are absorbed. In these events both gamma rays have likely undergone a photoelectric interaction with the scintillator material and the resulting photoelectron has been absorbed in the scintillator;
\item  the vertical bands below the peaks correspond to events where the photon absorbed in $S_{0}$ releases its whole energy in the scintillator, while the photon absorbed in $S_{1}$ releases only a fraction of its energy in the scintillator. In this case the second photon has likely undergone a Compton scattering, with the electron being absorbed in the scintillator and the scattered photon escaping from the detector;
\item  the horizontal bands on the left side of the peaks can be interpreted in a similar way as in the previous case, with the inversion of the roles of $S_{0}$ and $S_{1}$. The rate of these events is smaller than that in the previous region due to the smaller size of $S_{1}$ with respect to $S_{0}$;
\item the diagonal bands centered on the lines $E_{0}+E_{1}=1.17\units{MeV}$ and $E_{0}+E_{1}=1.33\units{MeV}$ correspond to events where one of the two photons undergoes Compton scattering in a scintillator, with the scattered photon being absorbed in the other scintillator;
\item the diagonal band centered on the line $E_{0}+E_{1}=2.50\units{MeV}$ corresponds to events where the first photon is absorbed in one of the two scintillators, the second photon undergoes Compton scattering in the same scintillator and the scattered photon is absorbed in the other scintillator;
\item the diagonal bands centered on the lines $E_{0}+E_{1}=1.46\units{MeV}$ and $E_{0}+E_{1}=2.61\units{MeV}$ correspond to events where a gamma-ray from the decay of $^{40}$K or $^{208}$Tl undergoes Compton scattering in a scintillator with the scattered photon being absorbed in the other scintillator. These decays mostly originate from the $^{40}$K in the glass windows of the PMTs and from the $^{208}$Tl in the scintillators.
\item the region $E_{0}>2.5\units{MeV}$ which is populated by events where either both the photons from the $^{60}$Co decays or the gamma-ray from the $^{208}$Tl decay are fully absorbed in $S_{0}$, in coincidence with a cosmic-ray event in $S_{1}$.
\end{itemize}

The same features can be observed also in Fig.~\ref{fig:signal1dspectra}, which shows the average rates as a function of the energies $E_{0}$ and $E_{1}$ and of their sum. Looking at the energy spectra in $S_{0}$ and $S_{1}$ we see that, as expected, the Compton shoulder in $S_{1}$ is more relevant than in $S_{0}$, due to its smaller size. Looking at the distribution of total energy deposited in the scintillators we also see that the rate of events in the peak corresponding to the $^{40}$K line is similar to that of events in the peak corresponding to the detection of both gamma rays from $^{60}$Co.  

\begin{figure}[!t]
\centering
\includegraphics[width=0.95\columnwidth]{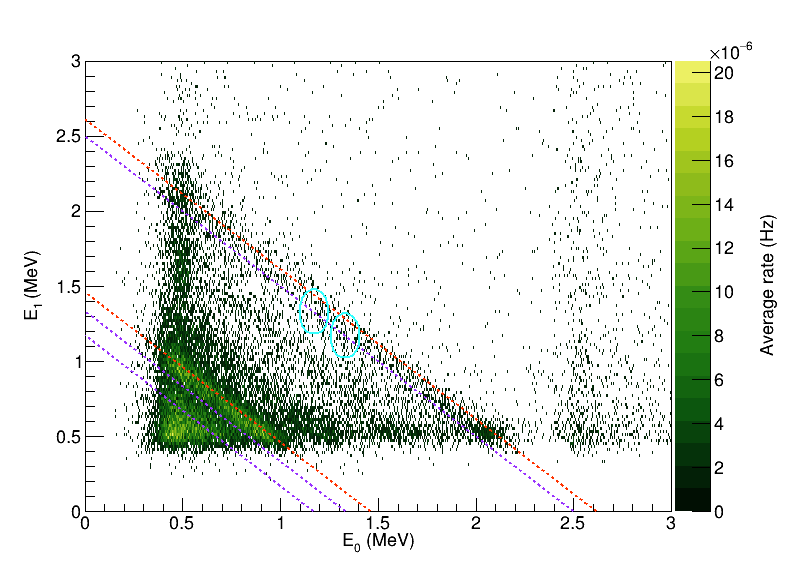}
\caption{Average rate of events as a function of the energy deposited in the two scintillators in the runs without the $^{60}$Co source. The lines have the same meanings as in Fig.\ref{fig:signal2dspectra}.  
}
\label{fig:bkg2dspectra}
\end{figure}

\begin{figure}[!t]
\centering
\includegraphics[width=0.95\columnwidth]{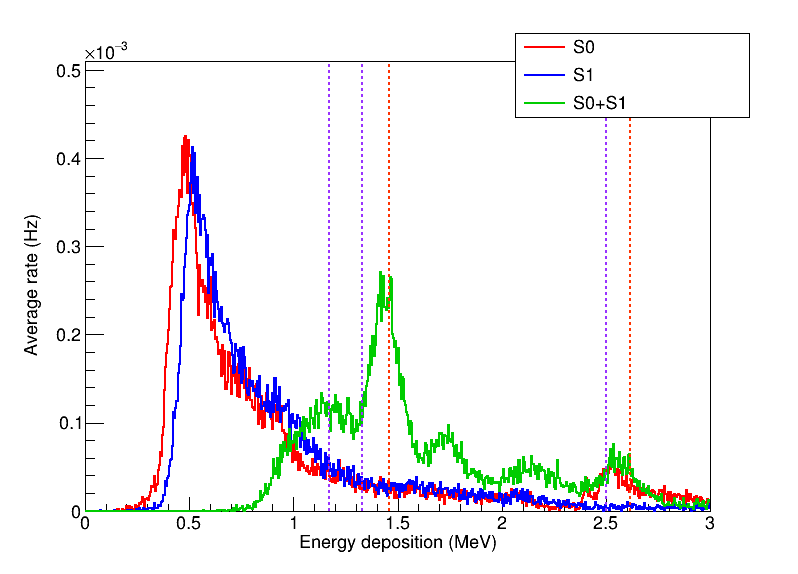}
\caption{Average rate of events as a function of the energy deposited in each of the two scintillators in the runs without the $^{60}$Co source. The lines have the same meanings as in Fig.~\ref{fig:signal1dspectra}.
}
\label{fig:bkg1dspectra}
\end{figure}

Figs.~\ref{fig:bkg2dspectra} and ~\ref{fig:bkg1dspectra} show the distributions of energy depositions in the two scintillators  obtained in the runs without the $^{60}$Co source. Comparing Fig.~\ref{fig:bkg2dspectra} with Fig.~\ref{fig:signal2dspectra}, we see a significant rate drop in the regions corresponding to events from $^{60}$Co. The diagonal bands corresponding to photons from $^{40}$K and $^{208}$Tl decays become more evident, and additional diagonal bands corresponding to photons produced in the decays of other radioactive isotopes are also visible. Looking at the shape of individual energy spectra of $S_{0}$ and $S_{1}$ shown in Fig.~\ref{fig:bkg1dspectra}, we see that they are dominated by cosmic-ray events, with a contribution from natural radioactivity. This contribution becomes more evident when looking at the distribution of the total energy deposited in the two scintillators.

\begin{figure}[!t]
\centering
\includegraphics[width=0.80\columnwidth]{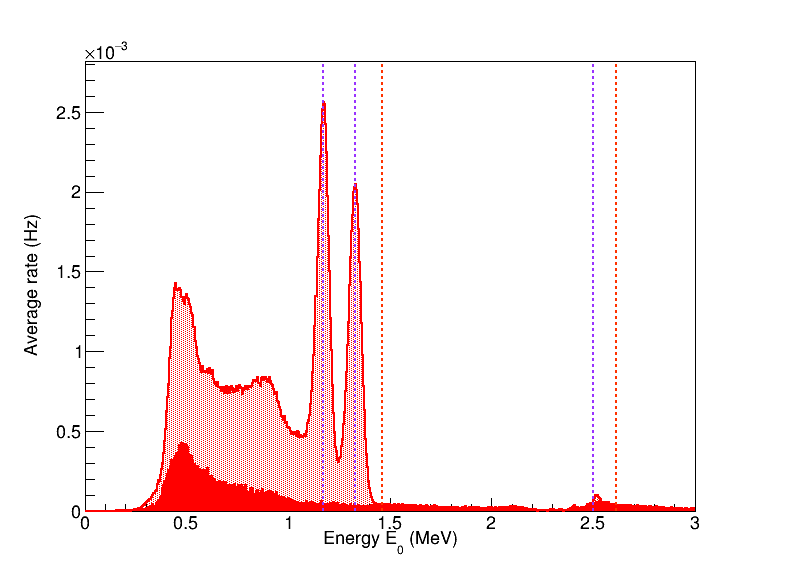}
\includegraphics[width=0.80\columnwidth]{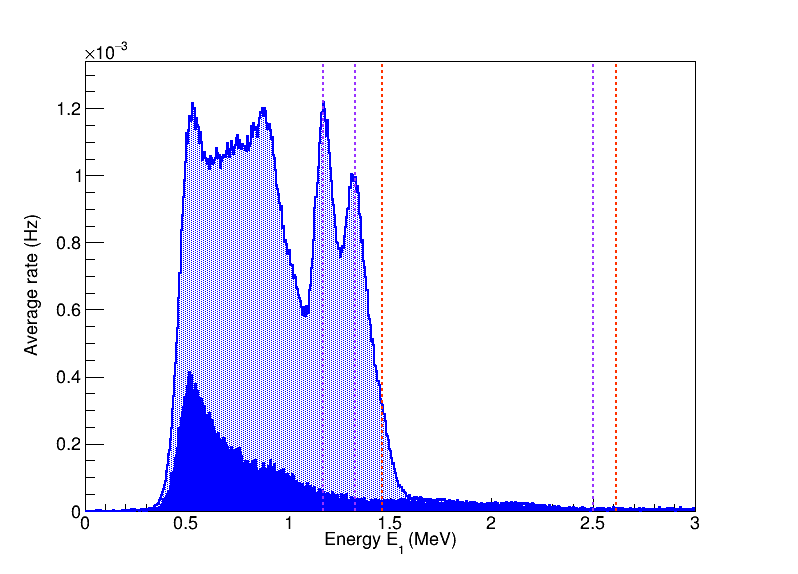}
\caption{Average rate of events as a function of the energy deposited in $S_{0}$ (top panel) and $S_{1}$ (bottom panel). The data collected in the runs with the $^{60}$Co source (histograms with shaded areas) are compared with those collected in the runs without the source (histograms with filled area). The lines have the same meanings as in Fig.~\ref{fig:signal1dspectra}.
}
\label{fig:comparison}
\end{figure}

A further comparison between the data taken in the runs with and without the $^{60}$Co source is shown in Fig.~\ref{fig:comparison}, where the energy spectra in the two scintillators are represented. We see that, for energy depositions above $1.33\units{MeV}$ in both scintillators, the rate of events with the $^{60}$Co source is consistent with the rate of events without the source. 

In our analysis we selected events in the regions corresponding to the two peaks shown in Fig.~\ref{fig:signal2dspectra}. The selection was performed discarding events outside the ellipses defined by the following equations:
\begin{eqnarray}
\frac{(E_{0}-1.17\units{MeV})^2}{a^2}  + \frac{(E_{1}-1.33\units{MeV})^2}{b^2} & = & 1 
\label{eq:ellipses1}
\\ 
\frac{(E_{0}-1.33\units{MeV})^2}{a^2}  + \frac{(E_{1}-1.17\units{MeV})^2}{b^2} & = & 1
\label{eq:ellipses2}
\end{eqnarray}
where $a=0.075\units{MeV}$ and $b=0.150\units{MeV}$. We set the values of $a$ and $b$ to about $3$ times the energy resolutions of $S_{0}$ and $S_{1}$ at $1\units{MeV}$. The contours of the ellipses in Eqs.~\ref{eq:ellipses1} and ~\ref{eq:ellipses2} are shown in the plots of Figs.~\ref{fig:signal2dspectra} and ~\ref{fig:bkg2dspectra}.

\newpage

\section{Results and discussion}
\label{sec:results}

\begin{figure}[!t]
\centering
\includegraphics[width=0.95\columnwidth]{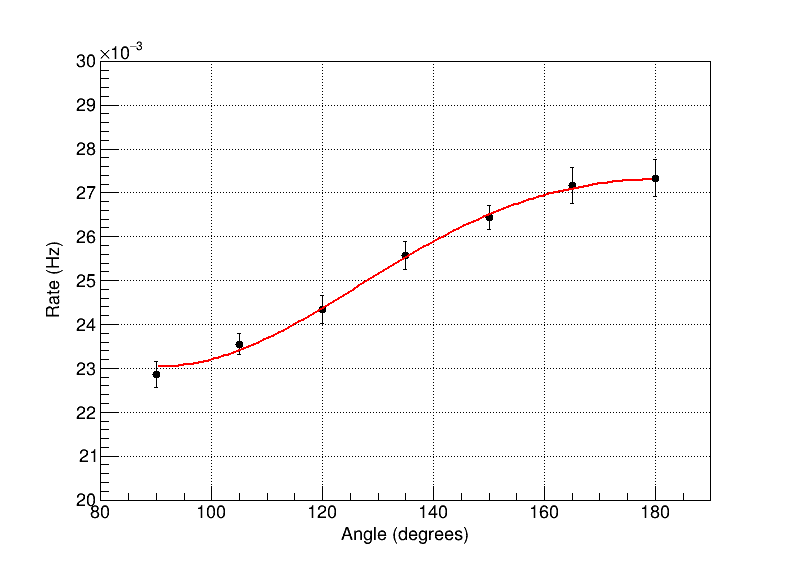}
\caption{Average rate of signal events as a function of the opening angle between $S_{0}$ and $S_{1}$. The red line represents the fit function.
}
\label{fig:ratevsangle}
\end{figure}

We performed several measurements, changing the opening angle between the two scintillators in the range between $90\degrees$ and $180\degrees$ with $15\degrees$ steps. The event selection illustrated in Sec.~\ref{sec:dataanalysis} was applied to the runs taken in the different configurations explored. The signal rate at each angle was evaluated as the difference between the rate of selected events with the $^{60}$Co source and the rate of selected events without the source. Our results are summarized in Fig.~\ref{fig:ratevsangle}. 

We have fitted the experimental data with the function:
\begin{equation}
r(\theta) = r_{0} \left( 1 + a \cos^{2}\theta + b \cos^{4} \theta \right) 
\end{equation}
where $r_{0}$ corresponds to the rate when the angle between the two scintillators is $90\degrees$ and the expected values of the parameters $a$ and $b$, according to Hamilton's theoretical model, are $0.125$ and $0.042$ respectively. In our fit we find the values $r_{0}=(2.303 \pm 0.021) \times 10^{-2} \units{Hz}$, $a=0.249 \pm 0.058$ and $b=(-6.43 \pm 5.88) \times 10^{-2}$ with a $\chi^{2}/d.o.f.=0.77/4$. 

As expected, we find that the signal rate increases when increasing the opening angle between the two scintillators. However, our results slightly deviate from Hamilton's model predictions. In particular, we find a discrepancy of about $2\sigma$ on both the parameters $a$ and $b$  with respect to their theoretical values. These discrepancies can be due to several reasons. In our setup we used a low-activity source and we had to place the scintillators very close to the source to keep a sufficiently high event rate during data taking. As discussed in Sec.~\ref{sec:setup}, this configuration implies an uncertainty on the opening angle which, on the basis only of pure geometrical considerations, is of about $16\degrees$. Placing the scintillators at larger distances from the source would allow the minimization of this uncertainty. Another possible cause could be the finite size of the $^{60}$Co source used in our measurement. In Hamilton's model it is assumed that the radioactive source is pointlike. However, as discussed in Sec.~\ref{sec:setup}, our source was encapsulated in a disk of $1\units{cm}$ diameter and $1\units{mm}$ thickness. When the mobile scintillator was rotated at different angles (see Fig.~\ref{fig:setup}), the source was always kept in the same position, and therefore its projected cross section on the front area of the mobile scintillator changed with the rotation angle.   

The measurement illustrated in this work can be improved in several ways. First of all, one could use two large size scintillators (i.e. at least with the size of $S_{0}$ in our setup) to enhance the probability that both gamma rays deposit their whole energy within the scintillator volumes. A further enhancement could be also obtained using a source with a higher activity, which would allow the scintillators to be placed at larger distances, thus improving the precision in the angle definition. Finally, the effects of possible asymmetries in the shape of the source could be minimized using a rotating source.

\section*{References}
\bibliographystyle{unsrt}
\bibliography{cobalt60.bib}{}

\end{document}